\documentstyle{elsart}     
\begin{document}

\renewcommand{\theequation}{\thesection.\arabic{equation}}

\def\e{\mbox{e}}
\def\ib{\,\mbox{i}\,}
\def\is{\,\mbox{\scriptsize i}\,}
\def\th{\hbox{th}}
\def\te{\vartheta_1}
\def\tv{\vartheta_4}
\def\s{\mbox{sh}}
\def\c{\mbox{ch}}
\def\case#1#2{{\textstyle{#1\over #2}}}

\begin{frontmatter}

\title{Excitations in the dilute $A_L$ lattice model:\\ 
$E_6$, $E_7$ and $E_8$ mass spectra}

\author[Canberra]{M.T. Batchelor} and
\author[LaTrobe]{K.A. Seaton}
\address[Canberra]{Department of Mathematics,
School of Mathematical Sciences,
Australian National University, Canberra ACT 0200, Australia}
\address[LaTrobe]{School of Mathematics, La Trobe University,
                 Bundoora VIC 3083, Australia}

\begin{abstract}
On the basis of features observed in the exact perturbation approach solution
for the eigenspectrum of the dilute $A_3$ model, we propose
expressions for excitations in the dilute $A_4$ and $A_6$ models. Principally, we require
that these expressions satisfy the appropriate inversion relations. We
demonstrate that they give the expected $E_7$ and $E_6$
mass spectra, and universal amplitudes, and agree with numerical
expressions for the eigenvalues. 
\end{abstract}

\end{frontmatter}

\section{Introduction}

\setcounter{equation}{0}

The dilute $A_L$ model is an exactly solvable, 
restricted solid-on-solid model defined on the square lattice. 
At criticality, the model can be constructed \cite{WNS,R} 
from the dilute $O(n)$ loop model \cite{N,WN}.
Each site of the lattice can take one of $L$ possible
(height) values, subject to the restriction that
neighbouring sites of the lattice either have the
same height, or differ by $\pm 1$.
Most importantly, the model can also be solved away
from criticality.
The off-critical Boltzmann weights of the allowed height
configurations of an elementary face of the lattice are
parametrised in terms of elliptic theta functions \cite{WNS}.
The interpretation of the elliptic nome $p$ differs according
to whether $L$ is even or odd. In particular, for $L$ odd 
the up-down symmetry of the Boltzmann weights is broken away 
from criticality. For
$L=3$ the elliptic nome plays the role of magnetic field.
Moreover, the dilute $A_3$ model provides, in one of its regimes, 
an integrable lattice 
realisation of the $E_8$ Ising model, being in the same
universality class as the two-dimensional Ising model in
a magnetic field.  

The calculation of the various off-critical thermodynamic 
properties of the model have verified this correspondence.
The singular part of the bulk free energy of the dilute $A_3$ model 
in the appropriate regime gives the magnetic Ising exponent
$\delta=15$ \cite{WNS}, which also follows from the calculation
of the local height probability \cite{WPSN}.
The expected Ising magnetic surface exponent $\delta_s = -\case{15}{7}$
follows from the excess surface free energy \cite{BFZ}.
Moreover the $E_8$ mass spectrum,
\begin{equation}
\begin{array}{ll}
m_2 = 2 \cos \frac{\pi}{5}  & = 1.618~033\ldots \\
m_3 = 2 \cos \frac{\pi}{30} & = 1.989~043\ldots \\
m_4 = 4 \cos \frac{\pi}{5} \cos \frac{7\pi}{30} & = 2.404~867\ldots \\
m_5 = 4 \cos \frac{\pi}{5} \cos \frac{2\pi}{15} & = 2.956~295\ldots \\
m_6 = 4 \cos \frac{\pi}{5} \cos \frac{\pi}{30}  & = 3.218~340\ldots \\
m_7 = 8 \cos^2 \frac{\pi}{5} \cos \frac{7\pi}{30} & = 3.891~156\ldots \\
m_8 = 8 \cos^2 \frac{\pi}{5} \cos \frac{2\pi}{15} & = 4.783~386\ldots
\end{array}
\label{masses}
\end{equation}
predicted by Zamolodchikov \cite{Za,Zb} for the Ising model in a
magnetic field is seen in the single particle 
excitation spectrum \cite{BNW,GN,MO,BSb}. Here the masses are
normalized such that $m_1=1$. They coincide with the components
of the Perron-Frobenius vector of the Cartan matrix of the Lie
algebra $E_8$. 
 
In this paper we consider off-critical excitations in the dilute $A_4$
and $A_6$ models, which are expected to be related to the $E_7$ and
$E_6$ scattering theories. 
The $E_6$ masses are (see, e.g., \cite{KM,BCD,F} and refs therein) 
\begin{equation}
\begin{array}{ll}
m_1 = m_{\bar 1} = 1 &\\ 
m_2 = 2 \cos \frac{\pi}{4}&=1.414~213\ldots\\ 
m_3 = m_{\bar 3} = 2 \cos \frac{\pi}{12}&= 1.931~851\ldots\\
m_4 = 4 \cos \frac{\pi}{4} \cos \frac{\pi}{12}& = 2.732~050\ldots 
\end{array}
\label{masses6}
\end{equation}
The $E_7$ masses, with $m_1=1$, are \cite{KM,BCD,F}
\begin{equation}
\begin{array}{ll}
m_2 = 2 \cos \frac{5\pi}{18}  & = 1.285~575\ldots \\
m_3 = 2 \cos \frac{\pi}{9} & = 1.879~385\ldots \\
m_4 = 2 \cos \frac{\pi}{18} & = 1.969~615\ldots \\
m_5 = 4 \cos \frac{\pi}{18} \cos \frac{5\pi}{18} & = 2.532~088\ldots \\
m_6 = 4 \cos \frac{\pi}{9} \cos \frac{2\pi}{9}  & = 2.879~385\ldots \\
m_7 = 4 \cos \frac{\pi}{18} \cos \frac{\pi}{9} & = 3.701~666\ldots \\
\end{array}
\label{masses7}
\end{equation}

Our approach begins in the next two sections by considering the 
inversion relations that hold for the off-critical dilute $A_L$ 
models, how our solution \cite{BSb} satisfies them in the case $L=3$, 
and how the $E_8$ structure manifests itself within the solution. 
In the subsequent sections we propose solutions for $A_4$ and 
$A_6$ and demonstrate the expected $E_7$ and $E_6$ mass spectra. 
We conclude with some numerical evidence and discussion.

\section{Inversion relations}

\setcounter{equation}{0}

The eigenvalues of the row transfer matrix of the dilute $A_L$ model, 
defined on a periodic strip of width $N$, where we take $N$ even,
are \cite{BNW}
\begin{eqnarray}
\Lambda(u) &=& \omega \left[
\frac{\te(2\lambda-u)\;\te(3\lambda-u)}{\te(2\lambda)\;\te(3\lambda)}
\right]^N
\prod_{j=1}^N
\frac{\te(u-u_j+\lambda)}{\te(u-u_j-\lambda)}
\nonumber \\
&&+ \left[
\frac{\te(u)\;\te(3\lambda-u)}{\te(2\lambda)\;\te(3\lambda)}
\right]^N
\prod_{j=1}^N
\frac{\te(u-u_j) \; \te(u-u_j-3\lambda)}
     {\te(u-u_j-\lambda) \; \te(u-u_j-2 \lambda)}
\nonumber\\
&& + \, \omega^{-1}
\left[
\frac{\te(u)\;\te(\lambda-u)}{\te(2\lambda)\;\te(3\lambda)}
\right]^N
\prod_{j=1}^N
\frac{\te(u-u_j-4\lambda)}{\te(u-u_j-2\lambda)} ,
\label{eigs}
\end{eqnarray}
where the $N$ roots $u_j$ are given by the Bethe equations
\begin{equation}
\omega \left[
\frac{\te(\lambda-u_j)}{\te(\lambda+u_j)}\right]^{N} =
-\prod_{k=1}^{N}
\frac{\te(u_j - u_k - 2\lambda) \; \te(u_j - u_k + \lambda) }
     {\te(u_j - u_k + 2\lambda) \; \te(u_j - u_k - \lambda) },
\label{BAE}
\end{equation}
with $\omega=\exp(\ib \pi \ell/(L+1))$ for $\ell=1,\ldots,L$.
For regime 2, the regime to be considered, the spectral parameter $u$
lies in the range $0<u< 3\lambda$, with $\lambda=\pi s /r$, where
$s = L+2$ and $r=4(L+1)$. 

The standard elliptic theta functions 
$\vartheta_1({u})$, $\vartheta_4({u})$ of nome $p$
are defined as
\begin{eqnarray}
\vartheta_1(u)&=&\vartheta_1(u,p)=2p^{1/4}\sin u\:
  \prod_{n=1}^{\infty} \left(1-2p^{2n}\cos
   2u+p^{4n}\right)\left(1-p^{2n}\right) , \label{theta1}  \\
\vartheta_4(u)&=&\vartheta_4(u,p)=\prod_{n=1}^{\infty}\left(
 1-2p^{2n-1}\cos2u+p^{4n-2}\right)\left(1-p^{2n}\right). 
  \label{theta4}
\end{eqnarray}

Also of use are the conjugate variables
\begin{equation}
w = \e^{-2\pi u/\epsilon} \quad \mbox{and} \quad 
x = \e^{- \pi^2/r \epsilon},
\end{equation}  
where nome $p=\e^{-\epsilon}$.
The relevant conjugate modulus transformations are
\begin{eqnarray}
\te(u,p) &=& \left( {\pi \over \epsilon} \right)^{1/2} 
             \e^{-(u-\pi/2)^2/\epsilon} \,
              E(w,q^2) ,  \\
\tv(u,p) &=& \left( {\pi \over \epsilon} \right)^{1/2} 
             \e^{-(u-\pi/2)^2/\epsilon} \,
              E(-w,q^2) , \label{conj}
\end{eqnarray}
where $q=\e^{-\pi^2/\epsilon}$ and 
\begin{equation}
E(z,p) = \prod_{n=1}^{\infty} (1-p^{n-1} z)(1-p^n z^{-1})(1-p^n).
\end{equation}

{}For this model, the partition function per site $\kappa$
was first calculated using the inversion relation \cite{WNS,WPSN}
\begin{equation}
\kappa(u) \, \kappa(u+3\lambda) = 
\frac{\te(2\lambda-u) \te(3\lambda-u) \te(2\lambda+u) \te(3\lambda+u)}
{\te^2(2\lambda) \te^2(3\lambda)}.
\label{ir1}
\end{equation}
In this way the bulk free energy per site
$f=\log \kappa$ was found to be 
\begin{equation}
f = \sum_{k=1}^{\infty}
\frac{(1-w^k)(1-x^{6sk}w^{-k})(x^{4sk}+x^{(2r-6s)k})(1+x^{2sk})}
{k(1-x^{2rk})(1+x^{6sk})}.
\label{fen}
\end{equation}
The same result was derived \cite{BSb,BS} from the Bethe Ansatz solution
for the groundstate eigenvalue $\Lambda_0(u)$.

Making use of the Poisson summation formula in the free energy (\ref{fen})
the leading singularity as  $p \to 0$ in regime 2 is
\begin{equation}
f\sim {\cal A} \, p^{r/3s},\label{fen2}
\end{equation}
where the amplitude ${\cal A}$ is given in terms of $L$ by
\begin{equation}
{\cal
A}=4\sqrt{3} \, \frac{\cos\left(\frac{\pi(L+6)}{6(L+2)}\right)}
{\sin\left(\frac{2\pi(L+1)}{3(L+2)} \right)},
\label{fenamp}
\end{equation}
and we have taken the isotropic value $u=3 \lambda /2$.

Excitations in the eigenspectrum can be considered in terms of the quantity
\begin{equation}
r_j(u) = \lim_{N \to \infty} \frac{\Lambda_j(u)}{\Lambda_0(u)}.
\end{equation}
The inversion relation (\ref{ir1}) is simply 
\begin{equation}
r_j(u) \, r_j(u+3\lambda) = 1, \label{ir2}
\end{equation}
but there is a further relation to be satisfied \cite{BSb}, 
\begin{equation}
r_j(u) \, r_j(u+2\lambda)=r_j(u+\lambda).\label{ir3}
\end{equation}

Our approach here is not to solve the inversion relations directly,
as was done, e.g., by Kl\"umper and Zittartz for the 
excitation spectra of the eight-vertex model \cite{KZ}.
Rather, in the light of our results for the excitations of
the dilute $A_3$ model, we use the above inversion relations to
give constraints on the Lie algebraic properties of a
conjectured solution. We then test our results 
as best we can by numerically diagonalising the transfer matrix, and
by comparison with results for $E_7$ and $E_6$ obtained by other methods.

\section{The dilute $A_3$ model and the $E_8$ mass spectrum}
 
\setcounter{equation}{0}

We now summarise our results \cite{BSb} for the dilute $A_3$ model,
obtained by the exact perturbation
approach \cite{Baxter}. The leading excitations in a given band of eigenvalues
can be written in the compact form
\begin{equation}
r_j(w) = w^{n(a)} \prod_a \frac{E(-x^a/w,\,x^{60}) E(-x^{30-a}/ w,\,x^{60})}
                                {E(-x^a w,\,x^{60}) E(-x^{30-a}w,\,x^{60})},
\label{compact}
\end{equation}
where the numbers $a$ and $n(a)$ are given in Table 1. 
The $E_8$ numbers $a$ have been discussed by McCoy and Orrick
for the related Hamiltonian \cite{MO}. They appear, e.g., 
in $E_8$ scattering theory \cite{BCD} and in $E_8$ Lie 
algebraic polynomials \cite{K}.   
The number $n(a)$ denotes the relevant band of eigenvalues. \\

\begin{table}[t]
\caption{
Parameters appearing in the eigenvalue expression (\ref{compact}). 
\vskip 5mm
}
%\centerline{
\begin{tabular}{||c|c|l||}
\hline
$j$ &  $n(a)$ & $a$ \\
\hline
1   &       2 & 1, 11 \\
2   &       2 & 7, 13 \\
3   &       3 & 2, 10, 12 \\
4   &       3 & 6, 10, 14 \\
5   &       4 & 3, 9, 11, 13 \\
6   &       4 & 6, 8, 12, 14 \\
7   &       5 & 4, 8, 10, 12, 14\\
8   &       6 & 5, 7, 9, 11, 13, 15 \\
\hline
\end{tabular}
\vskip 5mm
%}
\end{table}

Note that within a band of eigenvalues there may be more than
one class of excitation. For example, in the
leading band of eigenvalues there are two, which arise from a 2-string
and a 4-string structure in the Bethe roots \cite{BNW,GN}. The expression 
(\ref{compact}) is the leading excitation for each class of eigenvalue.
The last excitation within a class is also given by (\ref{compact}),
but with positive argument in the elliptic functions.

In the original variables (\ref{compact}) reads 
\begin{equation}
r_j(u) = \prod_a 
\frac{\tv(\frac{a\pi}{60}-\frac{8 u}{15},p^{8/15}) 
      \tv(\frac{(30-a)\pi}{60}-\frac{8 u}{15},p^{8/15})} 
     {\tv(\frac{a\pi}{60}+\frac{8 u}{15},p^{8/15})
      \tv(\frac{(30-a)\pi}{60}+\frac{8 u}{15},p^{8/15}) }.
\label{compactu}
\end{equation}

The various correlation lengths follow as
\begin{equation}
\xi_j^{-1} = - \log r_j(u),
\end{equation}
where we take the relevant leading 
eigenvalue at the isotropic point $u = 3\lambda/2$, which for $L=3$ is 
$u = \frac{15 \pi}{32}$.  

The fundamental correlation lengths can thus be written 
\begin{eqnarray}
m_j = \xi_j^{-1} 
&=& \sum_a \log \frac{
\tv(\frac{a\pi}{60}+\frac{\pi}{4},p^{8/15})
\tv(\frac{(30-a)\pi}{60}+\frac{\pi}{4},p^{8/15})}
{\tv(\frac{a\pi}{60}-\frac{\pi}{4},p^{8/15})
\tv(\frac{(30-a)\pi}{60}-\frac{\pi}{4},p^{8/15})}
\nonumber \\
&=& 2 \sum_a \log \frac{
\tv(\frac{a\pi}{60}+\frac{\pi}{4},p^{8/15})}
{\tv(\frac{a\pi}{60}-\frac{\pi}{4},p^{8/15})} .
\label{massesp}
\end{eqnarray}
Expanding this expression in powers of $p$ gives
\begin{equation}
m_j \sim 8\, p^{8/15}  \sum_a \sin \case{a\pi}{30}
\quad \mbox{as} \quad p \to 0 \, .
\end{equation}
This is the formula obtained by McCoy and Orrick \cite{MO} 
for the Hamiltonian,
from which the $E_8$ masses in (\ref{masses}) are recovered by
virtue of trig identities.

In particular,
\begin{equation}
\xi_1^{-1} \sim 8\, p^{8/15}  (\sin \case{\pi}{30}+\sin \case{11\pi}{30})
=16 \sin \case{\pi}{5} 
\cos \case{\pi}{6}\,\, p^{8/15}. \label{cor1}
\end{equation}

We are now able to consider the universal magnetic Ising 
amplitude \cite{BS}. From (\ref{fen2}) and (\ref{fenamp}),
\begin{equation}
f \sim 4 \sqrt 3 \, \frac{\sin{\pi \over 5}}{\cos{\pi \over 30}}
                   \, p^{16/15}
\quad \mbox{as} \quad p \to 0 \, .
\end{equation}

Combining this with (\ref{cor1}) gives 
\begin{equation}
f \, \xi_1^2 = {1 \over 16 \sqrt 3 \sin{\pi \over 5} \cos{\pi \over 30}}
             = 0.061~728~589 \ldots \quad \mbox{as} \quad p \to 0 \, .
\label{uni}
\end{equation}
This is the result for the universal magnetic Ising amplitude obtained 
earlier by thermodynamic Bethe Ansatz calculations based on the $E_8$ 
scattering theory \cite{F} (see also Ref.~\cite{DM} in the context of 
the form-factor bootstrap approach).
Here it has been obtained from the lattice model.

{}From the outset, no assumptions were made on the $E_8$ structure
in the dilute $A_3$ model, both in our own calculations, and in 
the thermodynamic Bethe Ansatz calculations \cite{BNW,MO}.  
We now highlight a few of the $E_8$ features as a  
guide to our considerations of $E_7$ and $E_6$.

{}First, each $a$ value occurs in (\ref{compact}) together with
its complement in 30, the Coxeter number of $E_8$, so that no integer 
greater than 15 appears in the lists in Table 1. 

Second, the inversion relation
\begin{equation}
r_j(w) \, r_j(x^{30} w) = 1, \label{ir81}
\end{equation}
is satisfied by an expression of the form (\ref{compact})
for any $a$ value. However, the stronger relation
\begin{equation}
r_j(w) \, r_j(x^{20}w)=r_j(x^{10}w), \label{ir82}
\end{equation}
is satisfied if, within the set of integers, one finds not only $a$,
where $a=1,\ldots,9$, but also $a+10$, or equivalently its complement 
in 30, $20-a$, by virtue of the properties
\begin{equation} 
E(z,p)=E(p/z,p)=-zE(z^{-1},p).
\end{equation}
The integer $a=10$ may appear alone, because
the factor it contributes to $r_j(w)$ satisfies (\ref{ir82}) by itself.
From Table 1, the sets of integers found by the perturbative 
approach \cite{BSb} all have these features. 

{}Finally, we observe that the nome $p$ cancels in (\ref{uni}) because 
of the relationship between the power of $p$ occurring in $f$ and in 
$\xi_1$. Indeed, this combination defines the hyperscaling relation
between the corresponding critical exponents. 

\section{The dilute $A_4$ model and the $E_7$ mass spectrum}
\setcounter{equation}{0}

We now use our observations for $E_8$ to arrive at a conjecture  
(equivalent to (\ref{compact})) for the excitations of $E_7$. 

The free energy expression is, from (\ref{fen2}) and (\ref{fenamp}),
\begin{equation}
f \sim  \frac{2\sqrt 3}{\sin{5\pi \over 18}}
                   \, p^{10/9}
\quad \mbox{as} \quad p \to 0 \, .
\end{equation}
In order to obtain a finite expression from $f \,\xi_1^2$, we thus expect
\begin{equation}
\xi_1^{-1} \sim  p^{5/9}
\quad \mbox{as} \quad p \to 0 \, .
\end{equation}
This power of the nome must appear in the expression
equivalent to (\ref{compactu}) for $E_7$, and is thus related to the
one we must propose for  $r_j(w)$ by the conjugate
modulus transformation (\ref{conj}), namely
\begin{equation}
\e^{-5\epsilon/9}\rightarrow \e^{-18 \pi^2/5\epsilon}=x^{72},
\end{equation}
where for $L=4$, $x=e^{- \pi^2/20 \epsilon}$.

The inversion relation in conjugate modulus form is
\begin{equation}
r_j(w) \, r_j(x^{36}w) = 1. \label{ir71}
\end{equation} 
However, the Coxeter number for $E_7$ is 18, so that we expect to 
select our integers from $1, \ldots, 9$. We thus
propose that the excitations for $E_7$ take the form
\begin{equation}
r_j(w) = w^{n(a)} \prod_a \frac{E(-x^{2a}/w,\,x^{72}) 
E(-x^{36-2a}/ w,\,x^{72})}
       {E(-x^{2a} w,\,x^{72}) E(-x^{36-2a}w,\,x^{72})}.
\label{compact7}
\end{equation}
The additional relation which serves to constrain the possible $a$ values is
\begin{equation}
r_j(w) \, r_j(x^{24}w)=r_j(x^{12}w).\label{ir72}
\end{equation}
This condition is satisfied if, within a set of possible integers, 
$a$ appears together with $a+6$ or equivalently $12-a$, apart from 
$a=6$ whose contribution satisfies (\ref{ir72}) by itself.

Written in terms of the original variables the expression (\ref{compact7}) is 
\begin{equation}
r_j(u) = \prod_a 
\frac{\tv(\frac{a\pi}{36}-\frac{5 u}{9},p^{5/9}) 
      \tv(\frac{(18-a)\pi}{36}-\frac{5 u}{9},p^{5/9})} 
     {\tv(\frac{a\pi}{36}+\frac{5 u}{9},p^{5/9})
      \tv(\frac{(18-a)\pi}{36}+\frac{5 u}{9},p^{5/9}) }.
\label{compact7u}
\end{equation}
Taking the isotropic value $u=9 \pi/20$ we obtain 
\begin{equation}
m_j=\xi_j^{-1}=2\sum_{a}
\log \frac{
\tv(\frac{a\pi}{36}+\frac{\pi}{4},p^{5/9})}
{\tv(\frac{a\pi}{36}-\frac{\pi}{4},p^{5/9})} 
\end{equation}
for the masses, and so 
\begin{equation}
m_j \sim 8\, p^{5/9}  \sum_a \sin \case{a\pi}{18}
\quad \mbox{as} \quad p \to 0 \, .\label{7masses}
\end{equation}

We now turn to the sets of integers associated with $E_7$ in the context of
Lie algebraic polynomials \cite{K} which form the first six rows of the
third column of Table 2. Clearly these integers satisfy the constraints
described above as being
placed upon them by (\ref{ir72}). Together with the last row, they are also
to be found within the table given for $E_7$ scattering in \cite{BCD}. \\

\begin{table}[t]
\caption{
Parameters appearing in the eigenvalue expression (\ref{compact7}). 
\vskip 5mm
}
%\centerline{
\begin{tabular}{||c|c|l||}
\hline
$j$ & $n(a)$ & $a$ \\
\hline
1&      1 & 6 \\
2&     2  & 1, 7 \\
3&   2    & 4, 8 \\
4&    2   & 5, 7 \\
5&   3    & 2, 6, 8 \\
6&   3    & 4, 6, 8 \\
7& 4      & 3, 5, 7, 9 \\
\hline
\end{tabular}
\vskip 5mm
%}
\end{table}

Applying trig identities to the sum in (\ref{7masses}) with these sets 
of integers gives
\begin{eqnarray}
\sum_{a=6}\sin \case{a\pi}{18}&=&\sqrt{3}/2, \nonumber \\
\sum_{a=1,7} \sin \case{a\pi}{18}&=&\sqrt{3}\cos\case{5\pi}{18}, \nonumber \\
\sum_{a=4,8} \sin \case{a\pi}{18}&=&\sqrt{3}\cos\case{\pi}{9}, \nonumber  \\
\sum_{a=5,7} \sin \case{a\pi}{18}&=&\sqrt{3}\cos\case{\pi}{18},\\
\sum_{a=2,6,8} \sin \case{a\pi}{18}&=&2\sqrt{3}\cos\case{\pi}{18}
\cos\case{5\pi}{18}, \nonumber  \\
\sum_{a=4,6,8} \sin \case{a\pi}{18}&=&
2\sqrt{3}\cos\case{\pi}{9}\cos\case{2\pi}{9}, \nonumber \\
\sum_{a=3,5,7,9} \sin \case{a\pi}{18}&=&
2\sqrt{3}\cos\case{\pi}{18}\cos\case{\pi}{9}, \nonumber
\end{eqnarray}
which, apart from normalisation, correspond to $m_1, \ldots, m_7$ of 
(\ref{masses7}).\footnote{
There is another relationship between the $E_7$ mass ratios, 
the trigonometric expression of (\ref{7masses}) and integers in the table 
of \cite{BCD}. However, the one described here is necessary in the context 
of the solvable dilute $A_4$ model in order to satisfy its inversion 
relations.}  
As another piece of evidence for our identification of
$a=6$ with $m_1$, from which the others follow, we consider the 
amplitude 
\begin{equation}
f \, \xi_1^2 = {2 \sqrt 3 \over \sin{5\pi \over 18} }\cdot 
{1 \over (8 \sin{\pi \over 3})^2 }
             ={1 \over 8\sqrt{3} \cos{2\pi \over 9}}  
\quad \mbox{as} \quad p \to 0 \, .
\label{uni7}
\end{equation}
This is in agreement with the $E_7$ thermodynamic Bethe Ansatz 
result \cite{F}.

\section{The dilute $A_6$ model and the $E_6$ mass spectrum}

The free energy
expression for the dilute $A_6$ model is, again from (\ref{fen2}) 
and (\ref{fenamp}),
\begin{equation}
f \sim  \frac{2\sqrt 6}{\cos{\pi \over 12}}
                   \, p^{7/6}
\quad \mbox{as} \quad p \to 0 \, , 
\end{equation}
and so we expect
\begin{equation}
\xi_1^{-1} \sim  p^{7/12}
\quad \mbox{as} \quad p \to 0 \, .
\end{equation}
This power of the nome must appear in the expression
equivalent to (\ref{compactu}) for $E_6$, and thus prescribes the nome of 
the expression we propose for  $r_j(w)$, because in the conjugate
modulus transformation (\ref{conj}),
\begin{equation}
\e^{-7\epsilon/12}\rightarrow \e^{-24 \pi^2/7\epsilon}=x^{96},
\end{equation}
where in the case $L=6$, $x=e^{- \pi^2/28 \epsilon}$. 
The inversion relation in conjugate modulus form is
\begin{equation}
r_j(w) \, r_j(x^{48}w) = 1. \label{ir61}
\end{equation}
Finally, the Coxeter number for $E_6$ is 12, so that we expect to 
select our integers from $1, \ldots, 6$. We thus
propose that the excitations for $E_6$ take the form
\begin{equation}
r_j(w) = w^{n(a)} \prod_a \frac{E(-x^{4a}/w,\,x^{96}) E(-x^{48-4a}/ w,\,x^{96})}
                                {E(-x^{4a} w,\,x^{96}) E(-x^{48-4a}w,\,x^{96})}.
\label{compact6}
\end{equation}

The additional relation which serves to constrain the possible $a$ values is
\begin{equation}
r_j(w) \, r_j(x^{32}w)=r_j(x^{16}w).\label{ir62}
\end{equation}
Thus within any set of possible integers, $a$ must appear together 
with $a+4$ or equivalently $8-a$, apart from $a=4$ which satisfies 
(\ref{ir62}) by itself.
Written in terms of the original variables the expression (\ref{compact6}) is 
\begin{equation}
r_j(u) = \prod_a 
\frac{\tv(\frac{a\pi}{24}-\frac{7 u}{12},p^{7/12}) 
      \tv(\frac{(12-a)\pi}{24}-\frac{7 u}{12},p^{7/12})} 
     {\tv(\frac{a\pi}{24}+\frac{7 u}{12},p^{7/12})
      \tv(\frac{(12-a)\pi}{24}+\frac{7 u}{12},p^{7/12}) }.
\label{compact6u}
\end{equation}
Taking the isotropic value $u=3 \pi/7$ we obtain 
\begin{equation}
m_j=\xi_j^{-1}=2\sum_{a}
\log \frac{
\tv(\frac{a\pi}{24}+\frac{\pi}{4},p^{7/12})}
{\tv(\frac{a\pi}{24}-\frac{\pi}{4},p^{7/12})} 
\end{equation}
for the masses. Thus 
\begin{equation}
m_j \sim 8\, p^{7/12}  \sum_a \sin \case{a\pi}{12}
\quad \mbox{as} \quad p \to 0 \, .\label{6masses}
\end{equation}

The integers given in Table 3 satisfy the constraint
placed upon them by (\ref{ir62}). Apart from the entry for $j=4$, 
these integers are again to be found in \cite{K}, and they appear within
the table of \cite{BCD} for $E_6$.

\begin{table}[t]
\caption{
Parameters appearing in the eigenvalue expression (\ref{compact6}). 
\vskip 5mm
}
%\centerline{
\begin{tabular}{||c|c|l||}
\hline
$j$ & $n(a)$ & $a$ \\
\hline
1 ,$\bar 1$&      1 & 4 \\
2&     2  & 1, 5 \\
3, $\bar 3$&   2    & 3, 5 \\
4&    3   & 2, 4, 6 \\
\hline
\end{tabular}
\vskip 5mm
%}
\end{table}

Applying trig identities to the sum in (\ref{6masses}) with these 
sets of integers gives
\begin{eqnarray}
\sum_{a=4}\sin \case{a\pi}{12}&=&\sqrt{3}/2, \nonumber \\
\sum_{a=1,5} \sin \case{a\pi}{12}&=&\sqrt{3}/\sqrt{2}, \nonumber \\
\sum_{a=3,5} \sin \case{a\pi}{12}&=&\sqrt{3}\cos\case{\pi}{12},   \\
\sum_{a=2,4,6} \sin \case{a\pi}{12}&=&\sqrt{6}\cos\case{\pi}{12},\nonumber
\end{eqnarray}
which, apart from normalisation, correspond to $m_1, \ldots, m_4$ 
of (\ref{masses6}).  Our identification of
$a=4$ with $m_1$, gives the amplitude
\begin{equation}
f \, \xi_1^2 = {2 \sqrt 6 \over \cos \case{\pi}{12} }
\cdot {1 \over (4 \sqrt{3})^2 }
             ={1 \over 2\sqrt{3} (1+\sqrt{3})}  
\quad \mbox{as} \quad p \to 0 \, ,
\label{uni6}
\end{equation}
which is in agreement with the thermodynamic Bethe Ansatz result \cite{F}.

\section{Numerical evidence and discussion}

\setcounter{equation}{0}

Based on our result (\ref{compact}) for the eigenspectrum of the 
dilute $A_3$ lattice model in regime 2, and its resulting $E_8$ 
structure, we have proposed analogous formulae for the dilute 
$A_4$ and $A_6$ models under the assumption of corresponding 
$E_7$ and $E_6$ structures. Such correspondence is to be expected
on a number of grounds. For example, at criticality the central 
charges of the dilute $A_L$ models are known from the underlying 
loop model \cite{WNS}. In regime 2, $c=\case{7}{10}$ for the 
$A_4$ model and $c=\case{6}{7}$ for the $A_6$ model. These
are the same as the $E_7$ and $E_6$ values \cite{KM}. 

A number of considerations have motivated our final results. 
Our first input was the hyperscaling relation, $f \xi^2 = {\rm constant}$,
which constrains the power of the elliptic nome $p$ appearing
in the inverse correlation lengths. We found that the stronger
inversion relation (\ref{ir3}) constrains the set of integers $a$ appearing 
in the eigenvalue expressions. We took 
these numbers from the Lie algebraic polynomials tabulated
by Kostant \cite{K}. Our results produce the $E_6$ (\ref{masses6})
and $E_7$ (\ref{masses7}) masses in the critical limit $p \to 0$. 
However, the configuration of $a$'s for the
heaviest mass does not appear in the Kostant polynomials. We chose
that configuration to be consistent with the predicted $E_6$ and
$E_7$ mass spectra, and subsequently noted that it had been observed in
the context of scattering theory \cite{BCD}.   
Our identification of the $a$'s associated with the lightest masses
also gives the universal amplitudes (\ref{uni6}) and (\ref{uni7}) 
in agreement with the thermodynamic Bethe Ansatz results based on
the $E_6$ and $E_7$ algebras \cite{F}. 

We have performed a number of numerical tests on the eigenspectra
of the dilute $A_4$ and $A_6$ models to check our results. 
Specifically, we have diagonalised the periodic row-transfer matrix
for finite lattice sizes. Consider the dilute $A_6$ model
first. Here $\lambda = \case{2\pi}{7}$. The largest eigenvalue 
$\Lambda_0$ is 3-fold degenerate in the thermodynamic limit.
Successive numerical
estimates of the first few eigenvalue ratios $\Lambda_0/\Lambda_j$
at the isotropic point $u = 3\lambda/2$
are tabulated in Table 4 for the values $p=0.1$ and $p=0.3$.
Excellent agreement is seen with the expected result (\ref{compact6u}),
which reduces to
\begin{equation}
\frac{\Lambda_0}{\Lambda_j} = \prod_a
\left[ \frac{
\tv(\frac{a\pi}{24}+\frac{\pi}{4},p^{7/12})}
{\tv(\frac{a\pi}{24}-\frac{\pi}{4},p^{7/12})} \right]^2.
\label{expect6}
\end{equation}

\begin{table}[t]
\caption{ 
Numerical estimates with increasing system size $N$ of 
leading eigenvalue ratios in the dilute $A_6$ model at
$\lambda = \case{2\pi}{7}$. Also shown is the expected
exact result (\ref{expect6}) in the thermodynamic limit.
The corresponding values of $a$ are as given in Table 3.
\vskip 5mm
}
%\centerline{
\begin{tabular}{||c|c|l|l|l||}
\hline
  & $N$ & $\Lambda_0/\Lambda_1$ & $\Lambda_0/\Lambda_2$ &
$\Lambda_0/\Lambda_3$ \\
\hline
$p=0.1$ & 3 & 6.0279 & ~      & ~  \\
        & 4 & 6.7882 & ~      & ~  \\
        & 5 & 6.9281 & ~      & ~  \\
        & 6 & 6.9474 & 15.268 & ~  \\
        & 7 & 6.9501 & 15.511 & 41.05 \\
 & $\infty$ & 6.9505 & 15.590 & 42.44 \\
\hline
$p=0.3$ & 3 & 89.93047 & &  \\
        & 4 & 90.08438 & &  \\
        & 5 & 90.08605 & 652.6278 &  \\
        & 6 & 90.08607 & 652.7399 & 6434.75\\
        & 7 & 90.08607 & 652.7442 & 6436.87\\
 & $\infty$ & 90.08607 & 652.7444 & 6437.08 \\
\hline
\end{tabular}
\vskip 5mm
%}
\end{table}

The dilute $A_4$ model at $\lambda = \case{3\pi}{10}$ is more 
complicated. A detailed numerical study of the
Bethe Ansatz equations has revealed all seven masses \cite{GNtbp}. 
However, the eigenvalue spectrum is dependent on the sign of $p$. 
In this case, all of the $E_7$ masses are observed in the 
$p < 0$ regime (regime $2^-$). Only a subset is observed for
$ p > 0$ (regime $2^+$). Our numerical results for the first
few leading eigenvalues are shown in Table 5 for $p=-0.3$.
The eigenvalues $\Lambda_1$ and $\Lambda_3$ 
do not appear in the eigenspectrum for $p=0.3$. Clearly
there is excellent agreement with our result ({\ref{compact7u}),
which here simplifies to 
\begin{equation}
\frac{\Lambda_0}{\Lambda_j} = \prod_a
\left[ \frac{
\tv(\frac{a\pi}{36}+\frac{\pi}{4},p^{5/9})}
{\tv(\frac{a\pi}{36}-\frac{\pi}{4},p^{5/9})} \right]^2.
\label{expect7}
\end{equation}
We expect this result to hold in regime $2^-$ for all of the 
masses, or correspondingly for each set of $a$'s given in
Table 2. Apart from $\Lambda_1$ and $\Lambda_3$,
we have not explored further which of the eigenvalues 
are absent in regime $2^+$.
We await the publication of Ref. \cite{GNtbp}.

In contrast with the dilute $A_4$ model, the
mass spectrum of the dilute $A_6$ model
appears to be equivalent in regimes $2^{\pm}$.
Such equivalence holds for the dilute $A_L$ models
with $L$ odd, where the eigenspectrum is independent
of the sign of $p$. This is a consequence of the off-critical 
weights breaking the $Z_2$ symmetry for $L$ odd. However, for
$L$ even this symmetry is not broken. As to why the mass spectrum
may be the same for the dilute $A_6$ model in regimes $2^{\pm}$,
this remains one of the mysteries of the dilute $A_L$ models for
$L$ even, which are yet to be investigated.
 
{}Finally we note that although the evidence for our conjectured 
results is convincing, they of course await a formal derivation.

\begin{table}[t]
\caption{
Numerical estimates with increasing system size $N$ of
leading eigenvalue ratios in the dilute $A_4$ model at
$\lambda = \case{3\pi}{10}$. Also shown is the expected
exact result (\ref{expect7}) in the thermodynamic limit.
The corresponding values of $a$ are as given in Table 2.
\vskip 5mm
}
%\centerline{
\begin{tabular}{||r|c|l|l|l|l||}
\hline
  & $N$ & $\Lambda_0/\Lambda_1$ & $\Lambda_0/\Lambda_2$ &
$\Lambda_0/\Lambda_3$ & $\Lambda_0/\Lambda_4$ \\
\hline
$p=-0.3$ & 4 & 116.09490 & 492.5475 &  & \\
         & 5 & 116.09969 & 493.2263 & 8669.13 &  \\
         & 6 & 116.09973 & 493.2413 & 8724.17 & 11928\\
         & 7 & 116.09973 & 493.2416 & 8726.53 & 12067\\
 & $\infty$  & 116.09973 & 493.2416 & 8726.64 & 12190\\
\hline
\end{tabular}
\vskip 5mm
%}
\end{table}

\ack
It is a pleasure to thank Uwe Grimm, Bernard Nienhuis, Will Orrick,
Ole Warnaar and Yu-kui Zhou for some helpful remarks. This work was 
undertaken while KAS was a visitor in the Department of Mathematics  
at The Australian National University.
The work of MTB has been supported by the Australian Research Council.

\end{document}